# Defect theory under steady illuminations and applications

Guo-Jun Zhu[1,2], Yi-Bin Fang[1,2], Zhi-Guo Tao[1,2], Ji-Hui Yang[1,2*], and Xin-Gao Gong[1,2]

[1]*Key Laboratory for Computational Physical Sciences (MOE), State Key Laboratory of Surface Physics, Department of Physics, Fudan University, Shanghai 200433, China*
[2]*Shanghai Qizhi Institution, Shanghai 200232, China*

Email: jhyang04@fudan.edu.cn;

## Abstract

Illumination has been long known to affect semiconductor defect properties during either growth or operating process. Current theories of studying the illumination effects on defects usually have the assumption of unaffected formation energies of neutral defects as well as defect transition energy levels, and use the quasi-Fermi levels to describe behaviors of excess carriers with conclusions at variance. In this work, we first propose a method to simulate steady illumination conditions, based on which we demonstrate that formation energies of neutral defects and defect transition energy levels are insensitive to illumination. Then, we show that optical and thermal excitation of electrons can be seen equivalent with each other to reach a steady electron distribution in a homogeneous semiconductor. Consequently, the electron distribution can be characterized using just one effective temperature $T'$ and one universal Fermi level $E_F'$ for a homogeneous semiconductor under continuous and steady illuminations, which can be seen as a combination of quasi-equilibrium electron system with $T'$ and a lattice system with $T$. Using the new concepts, we uncover the universal mechanisms of illumination effects on charged defects by treating the band edge states explicitly in the same footing as the defect states. We find that the formation energies of band edge 'defect' states shift with increased $T'$ of electrons, thus affecting the $E_F'$, changing defect ionic probabilities, and affecting concentrations of charged defects. We apply our theory to study the illumination effects on the doping behaviors in GaN:Mg and CdTe:Sb, obtaining results in accordance with experimental observations. More interesting experimental defect-related phenomena under steady illuminations are expected to be understood from our theory.

# I. Introduction

Defects play central roles in determining various properties of semiconductors[1-4]. However, defect characterizations are very challenging experimentally. Thanks to the development of defect theories, especially the first-principles defect calculations, semiconductor devices and applications have been greatly advanced in the past decades [5-7]. Nevertheless, current theoretical defect studies mainly focus on semiconductors under equilibrium conditions without considering practical operations. Recently, illumination was reported to influence the device performance and material properties for photocatalytic and photovoltaic semiconductors by affecting defect behaviors during either operating or growth processes [8-12]. The underlying mechanisms, however, remain elusive. Defect theories under illuminations are therefore necessary to further promote the development of semiconductor techniques.

Among various defect-related properties, defect formation energy and transition energy level are the two most crucial quantities: the former determines the defect concentration, and the latter, defined as the energy cost to get ionized, determines the ability of a defect to provide carriers. How illumination plays roles in affecting these two quantities are hence the first problem to be solved. Several theoretical schemes have been proposed so far to study the illumination effects on defect formations by: 1) assuming illuminations do not change the formation energies of neutral defects and defect transition energy levels; 2) using the quasi-Fermi levels (QFLs) to define formation energies of charged defects [9, 13-17]. These schemes are unsatisfactory, nonetheless, because the former assumption is not justified due to lack of efficient calculation methods to simulate the illumination conditions for defect supercells, while the latter has the problem of defining QFLs when various defects, excess electrons and holes are present together under illuminations. In fact, different works using different definitions yield different results. For example, Alberi defined a QFL for each kind of defects and reported that the concentration of dominate defects increased while the concentration of compensational defects decreased under illumination [14]. In contrast, Cai et al. assumed two quasi Fermi reservoirs and proposed a weight to define different

QFLs for donors, acceptors, and free carriers. They found that all charged defects tend to have the increased formation energies under illumination for any semiconductors [9]. In addition, the definitions of defect QFLs in some works rely on the carrier capture and emission rates, which are often difficult to be obtained accurately both theoretically and experimentally[9, 14, 15]. To understand defect behaviors under illuminations, defect calculation methods considering illumination conditions are eager to be developed and universal defect theories under illuminations should be established thereafter.

In this work, we first propose a method to simulate continuous and steady illumination conditions, based on which we demonstrate that formation energies of neutral defects and defect transition energy levels are insensitive to illumination. Then, we show that optical and thermal excitation of electrons can be approximately seen as equivalent with each other to reach a steady electron distribution. Consequently, the electron distribution can be characterized using just one effective temperature $T'$ and one universal Fermi level $E_F'$. And a semiconductor under continuous and steady illuminations can be seen as a combination of quasi-equilibrium electron system with temperature $T'$ and an equilibrium lattice system with temperature $T$. Under illuminations, by treating the band edge states explicitly in the same footing as the defect states, the formation energies of band edge 'defect' states shift with increased $T'$ of electrons, thus affecting the $E_F'$ of the electron system, changing the ionic probabilities of defect states, and affecting concentrations of charged defects. We apply the present theory to study the illumination effects on the doping behaviors in GaN:Mg and CdTe:Sb and obtain consistent results with experimental observations. We expect that more interesting experimental defect-related phenomena under steady illuminations can be understood from our theory.

## II. Illumination effects on neutral defect

Under defect dilute approximations, illuminations mainly excite electrons from the valence bands to the conduction bands. With the help of phonons, the photogenerated electrons (holes) will soon be relaxed to the band edges and the whole system will reach a steady state, as shown in Fig. 1a. To simulate excess carriers under continuous and steady illuminations, previous works often adopted a method of electron-occupation-constraining scheme, i.e., by reducing the occupation number at the valance band maximum (VBM) state and increasing the occupation number at the conduction band minimum (CBM) state simultaneously [18]. However, when a defect is created in a semiconductor supercell, one cannot find the exact VBM or CBM states any more due to band couplings [19, 20].

To simulate defects under illuminations, here we adopt an alternative way to simulate excess carriers by constraining charge densities instead of electron occupations. As we know, illuminations change the electron occupations and thus the total charge density distributions. Therefore, if we use the correct charge density distributions under illuminations, according to the density functional theory (DFT) we can obtain the correct total energy of a defective supercell, which we mainly concern in defect calculations. To mimic the steady state under illuminations, we add a charge density correction, i.e., $\Delta\rho^\sigma(r) = \lambda[\rho^\sigma_{CBM}(r) - \rho^\sigma_{VBM}(r)]$ ($\lambda$ is used to represent the illumination strength, $\rho^\sigma_{CBM}(r)$ and $\rho^\sigma_{VBM}(r)$ are the partial charge density of the spin-polarized CBM and VBM states, respectively), to the total charge density without illuminations. Note that, for simulations of semiconductors without any defects, the charge correction method is exactly equivalent to the electron-occupation-constraining scheme. However, for defective supercells, our method is physically more meaningful [19]. We implement the charge correction method in the *Quantum Espresso* code[21] and the flowchart is shown in Fig. 1b.

Now we can calculate the formation energies of neutral defects under illuminations according to the definition, which is:

$$\Delta H_f^{il}(\alpha, 0) = E^{il}(\alpha, 0) - E^{il}(host) + \sum n_i(\mu_i + E(i)), \qquad (1)$$

where $E^{il}(\alpha, 0)$ and $E^{il}(host)$ are the total energies of the supercell under illuminations with a neutral defect α and without any defects, respectively. $E(i)$ refers to the total energy of the element $i$ in its pure stable phase, $\mu_i$ is the chemical potential of element $i$ referenced to $E(i)$ and $n_i$ is the number of $i$ atoms removed from (positive) or put in (negative) the supercell in the process of defect formation. The defect transition energy levels under illuminations $E_t^{\alpha,il}(0/q)$ can also be calculated as:

$$E_t^{\alpha,il}(0/q) = [E^{il}(\alpha,q) - E^{il}(\alpha,0)]/(-q) + \varepsilon_{VBM}, \quad (q < 0),$$
$$E_t^{\alpha,il}(0/q) = \varepsilon_{CBM} - [E^{il}(\alpha,q) - E^{il}(\alpha,0)]/(-q), \quad (q > 0), \quad (2)$$

where $E^{il}(\alpha,q)$ is total energy of the supercell under illuminations with a defect α charged $q$, $\varepsilon_{VBM}$ and $\varepsilon_{CBM}$ are the energy levels of the CBM and VBM states, respectively. Other defect properties under steady illuminations such as defect diffusion barriers and defect-assisted non-radiative recombination can also be processed in this scheme.

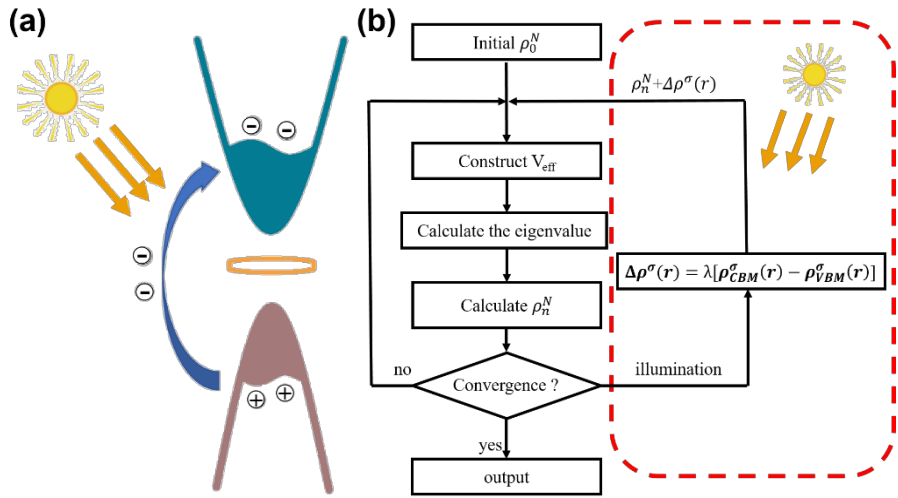

Figure 1. The simulation method of steady illumination. (a) The diagram to show the electron excitations and occupations under illuminations. (b) The sketching flow chart for the self-consistent calculations under steady illuminations.

We apply our method to study defect properties of MoS$_2$ monolayer and bulk, which is a representative system to show how illuminations affect defect properties in two-dimensional and three-dimensional materials. As an important semiconductor that has been studied abundantly, the defects in MoS$_2$ are essential in determining electrical [22, 23], magnetic[24], and optical properties[25, 26]. Our calculated formation energies of selected neutral defects in monolayer MoS$_2$ (see the Supplemental Materials for the defects in bulk MoS$_2$) including both intrinsic defects $V_S$, $V_{Mo}$, $Mo_i$, $S_i$ and the impurity $Nb_{Mo}$ as functions of illumination strengths are shown in Fig. 2a and the corresponding defect transition energy levels are given in Fig. 2b. Here the illumination strengths are corresponding to an excess carrier density of 0 and $6.4\times10^{12}$ cm$^{-2}$ for $\lambda = 0$ and 0.2, respectively. Our results without illuminations agree well with previous works (see discussion in the Supplemental Materials) [27, 28]. With the increase of illumination strengths, we find that the formation energies of neutral defects and defect transition energy levels have little changes. As shown in Fig. 2, the change of formation energy is less than 0.1 meV and that of defect transition energy level is less than 1 meV. Such phenomena can be attributed to the delocalization of band edge states, which have negligible effects on localized properties such as formation and ionization of an isolated defect. Our results demonstrate that illumination effects on formation energies of neutral defects and defect transition energy levels can be reasonably discarded, but other defect properties such as diffusion barriers should be dealt with carefully. In the followings, we will not distinguish formation energies of neutral defects and defect transition energy levels with or without illuminations unless otherwise stated.

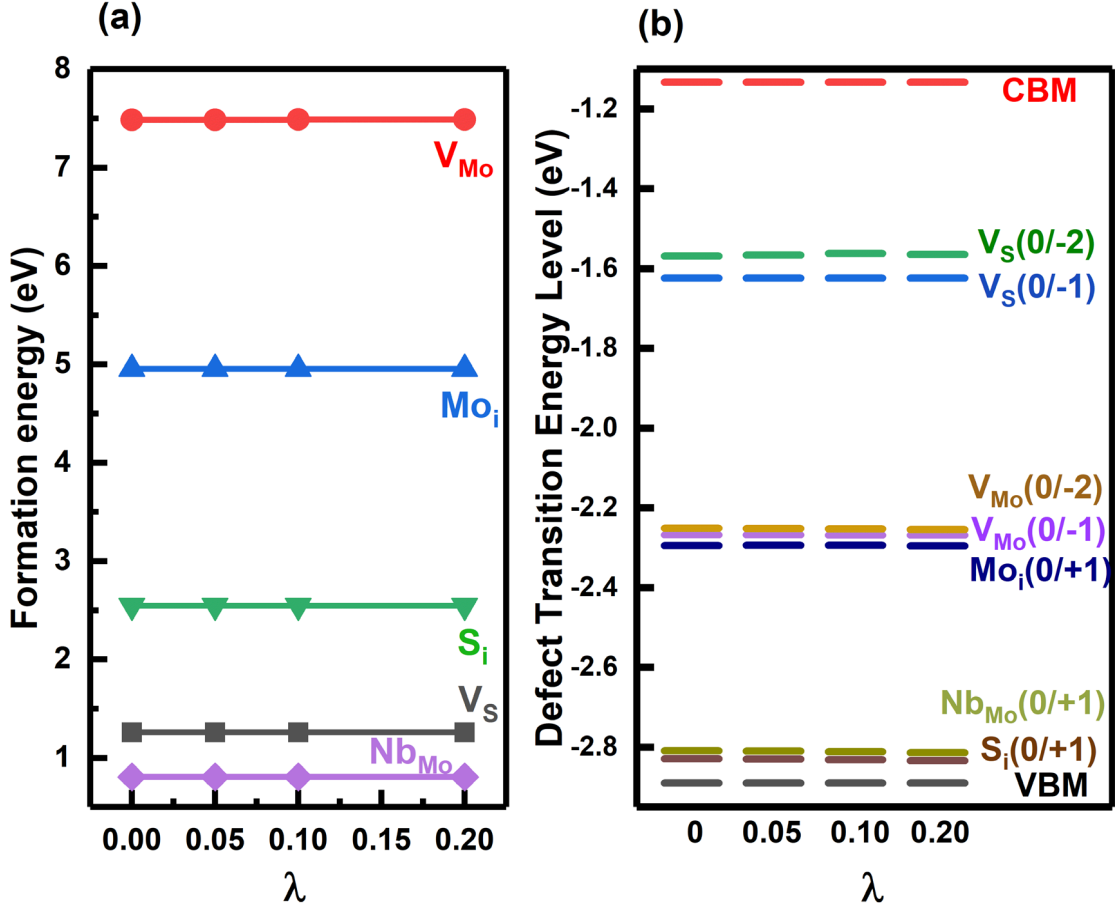

Figure 2. Illumination effects on formation energies of neutral defects and defect transition energy levels in MoS$_2$ monolayer. (a) Formation energies of neutral defects and (b) defect transition energy levels as functions of illumination strengths. The λ represents the number of excited electrons at the band edges in the supercell.

## III. Illumination effects on charged defect

Different from formation of neutral defects which is only related to crystal lattices and little affected by illuminations, formation of charged defects relied on electron potentials, that is, the Fermi reservoirs. How to define the Fermi level or QFLs under illuminations has been a key challenge in this field. We start to think of this problem from the excitation of electrons in a semiconductor with some defect α dilute and homogenously distributed, which is initially at a thermal equilibrium state with a lattice temperature of $T$, a Fermi level of $E_F$, and a free electron (hole) density of $n_0$ ($p_0$) (see Fig. 3a). We keep the total amount of defects fixed to focus solely on the electron behaviors at this stage as we will show the electron and lattice systems can be dealt with separately. We consider thermal excitations first. When the temperature is increased to $T'$, more carriers are generated from thermal excitations of both band and defect states. In the meanwhile, carriers are recombined via band-to-band transitions or via defect levels. When the generation rate is equal to the recombination rate, i.e.:

$$G = R_{BB} + R_{Aug} + \sum_{\alpha,q} R_{SRH}(\alpha, q), \qquad (3)$$

where $R_{BB}$, $R_{Aug}$, and $R_{SRH}(\alpha, q)$ are the band-to-band, Auger, and defect-assisted Shockley-Read-Hall recombination rates, respectively, electrons reach a steady distribution, that is, concentrations of free carriers, ratio of neutral and charged defects don't change any more, as shown in Fig. 3b. In this case, electrons reach an equilibrium state when the electron potentials in the conduction bands ($E_{Fc}$), valence bands ($E_{Fv}$), and defect states ($E_{FD}$) are the same, i.e., electrons in the whole system share the same Fermi level $E_F'$. From statistical physics, the electron potentials are defined according to:

$$n' = N_C(T') \exp\left(-\frac{E_C - E_{Fc}}{k_B T'}\right), \qquad (4)$$

$$p' = N_V(T') \exp\left(-\frac{E_{Fv} - E_V}{k_B T'}\right), \qquad (5)$$

$$N(\alpha, q) = N(\alpha, 0) exp\left(\frac{q * (E_t^\alpha - E_{FD})}{k_B T'}\right), \qquad (6)$$

$$E_{Fc} = E_{Fv} = E_{FD} = E_F'. \qquad (7)$$

Here, $N_C(T')$ and $N_V(T')$ are the effective density of states for the conduction band and the valence band at $T'$, respectively. $E_C$ and $E_V$ are the energy level of the CBM and VBM, respectively. $k_B$ is the Boltzmann constant, $N(\alpha, q)$ and $N(\alpha, 0)$ are the concentrations of charged and neutral defects, respectively. $E_{Fc}$, $E_{Fv}$, and $E_{FD}$ are the Fermi level of electrons at the conduction bands, valence bands, and defect states, respectively.

Now, we consider optical excitation (see Fig. 3c). Instead of increasing temperature to $T'$, we apply illumination to excite electrons in the system while keeping the lattice temperature $T$ unchanged (the heating effect of illumination on lattices is neglected due to, i.e., dissipations to the environment). In principle, the same electron distribution as in the case of Fig. 3b can be achieved by applying proper illuminations, especially if $T'$ does not differ too much from $T$ so that $N_C$ and $N_V$ don't change significantly before and after illuminations. This is reasonable because the behaviors of hot carriers after thermal or optical excitations are similar and one cannot distinguish thermal from optical excitations by just considering electron distributions. Because the two electron distributions are approximately equivalent, we can say that the electrons in Fig. 3c have an effective temperature of $T'$ and a Fermi level of $E_F'$. Note that, the different temperatures for electrons and lattices in Fig. 3c indicate that the system is not a fully equilibrium system. Instead, it can be regarded as a combination of a quasi-equilibrium electron system and an equilibrium lattice system.

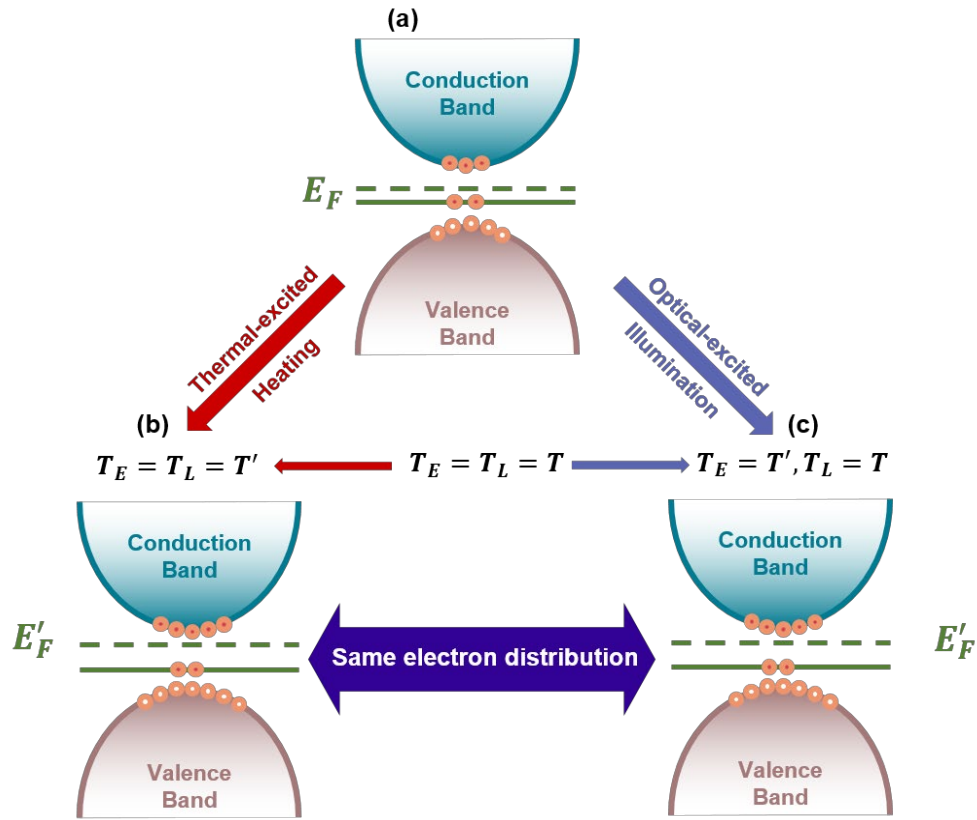

Figure 3. Diagrams to show the thermal and optical excitation processes of electrons in a semiconductor. (a) The band diagram to show the electron distributions in an equilibrium semiconductor system with defects under no illuminations. Note that, electrons have the same temperature with the lattice and a universal Fermi level of $E_F$. (b) The band diagram to show the electron distributions after reaching a new equilibrium state due to temperature increase from T to $T'$. Now the system has a universal Fermi level of $E_F'$. (c) The band diagram to show the electron

distributions after reaching a steady state under continuous illuminations. Note that, the electron distribution in Figs. b and c are equivalent. $T_E$ means effective electron temperature, $T_L$ means the lattice temperature.

From above discussion, we learn that thermal excitation of electrons is equivalent to optical excitation. Both thermal excitations and illuminations can 'heat' electrons while the lattice can only be heated thermally. Consequently, when a homogeneous semiconductor under continuous illuminations reaches a steady state with a certain distribution of electrons, we can define an effective temperature $T'$ and an effective Fermi level $E_F'$ according to Eqs. (4)-(7) to characterize the electron distribution. In practice, $T'$ can be self-consistently determined through Eqs. (3)-(7), given carrier generation rate due to illumination strengths as well carrier recombination rates. Alternatively, $T'$ can also be known according to Eqs. (4) and (5) if total carriers $n'$ and $p'$ are known. In any case, we can use $T'$ or $n'p'$ to describe the illumination strengths without considering complicated parameters like recombination rates. Meanwhile, a real temperature $T$ should be used to define properties unrelated to the Fermi reservoir, i.e., the equilibrium concentration of neutral defects under illuminations should be still determined according to:

$$N^{il}(\alpha, 0) = N(\alpha, 0) = gN_{site}(\alpha) \exp\left(\frac{-\Delta H_f(\alpha, 0)}{k_B T}\right), \qquad (8)$$

where $N_{site}(\alpha)$ is the number of possible sites of defect $\alpha$ in a supercell, and $T$ is the real lattice temperature, $g$ is the degeneracy factor of the electron occupations. For charged defects under illuminations, the equilibrium concentration should be determined according to Eq. (6) as it is related to the Fermi reservoir. In addition, the overall charge neutrality relation remains, that is,

$$\sum_{\alpha,q} q * N^{il}(\alpha, q) + p' - n' = 0. \qquad (9)$$

By self-consistently sovling Eqs. (3)-(9), we can obtain defect concentrations, carrier densities, and the effective Fermi level in a semiconductor at given illuminations (i.e., given generation rate G, $T'$ or $n'p'$).

Now we turn to study the illumination effects on formation of charged defects. Without loss of generality, we consider two kind of defects (A and B) and assume they are all at their ionized states ($A^+$ and $B^-$) with transition energy levels of $E_t^A$ and $E_t^B$. in real systems, defects could be vacancies, interstitials, antisites, or even complexes and have various charged states. Nevertheless, our analysis holds for any cases. Figs. 4a-4c show defect formation energies as functions of Fermi levels following conventional diagrams in defect calculations. It is noteworthy that as both formation energies of neutral defects and defect transition energy levels do not change with illumination strengths, the formation energy lines for charged defects [the solid lines, $\Delta H_f^{il}(A^+) = \Delta H_f^{il}(A^0) - q * (E_t^A - E_F')$ , $\Delta H_f^{il}(B^-) = \Delta H_f^{il}(B^0) - q * (E_t^B - E_F')$] do not change under illuminations either. Because illuminations mainly play roles through affecting the band edge excitations, we should treat the band edge states explicitly in the same footing as the defect states. As we did in Ref.[29], the electron occupation at the conduction band under illuminations (see Eq. (4)) can be treated as having a singly charged "acceptor" with its formation energy of $\Delta H_f^{il}(n') = k_B T' ln[\frac{gN_{site}}{N_C(T')}] + E_g - E_F'$ and a transition energy level at the VBM. Similarly, hole occupation in the valence band under illuminations (see Eq. 5) can be treated as an effective, singly ionized "donor" with its formation energy of $\Delta H_f^{il}(p') = k_B T' ln[\frac{gN_{site}}{N_V(T')}] + E_F'$ and a transition energy level at the CBM. The formation energies of band edge 'defects' are shown in dashed lines in Figs. 4a-4c.

Without illuminations, i.e., $T'=T$, the entanglements between defects and band edge excitations have been discussed in our previous work [29]. Here as an example, we just discuss the case of defect excitation dominant over thermal excitation under no illuminations, that is, the dashed lines are both above the solid lines in Fig. 4a. Other cases can be analyzed in a similar way. With the applied of illumination strengths, $T'$ increases and so do $N_V(T')$ and $N_C(T')$. Usually, the dashed lines for $\Delta H_f^{il}(n')$ and $\Delta H_f^{il}(p')$ will first shift upwards [due to $N_V(T'), N_C(T') < gN_{site}$] and then shift downwards. From Fig. 4a to Fig. 4b, the Fermi level has little changes as the $E_F'$ is always mainly determined by defects. However, due to the increase of $T'$, the ratio between charged and neutral defects will decrease according to Eq. (6). As the concentration of neutral defects don't change under illumination, the concentration of charged defects will decrease for both dominate and compensated defects.

When the illumination strengths are strong enough to make the band edge excitations comparable with defect excitations, i.e., $\Delta H_f^{il}(n')$ line is lower than $\Delta H_f^{il}(B^-)$ and/or $\Delta H_f^{il}(p')$ line is lower than $\Delta H_f^{il}(A^+)$, the $E_F'$ will be determined by both band edge and defect excitations. In this case, the $E_F'$ can move to the middle of the band gap or to the band edges, depending on practical situations, and the change of the concentration of charged defects has no definite direction but has to be determined according to Eq. (9) (more detail discusses are presented in Supplement Materials). When the illumination intensity is further increased so that the defect excitation is not important compared to band edge excitations (see Fig. 4c), the $E_F'$ is just determined by the $N_V$ and $N_C$. An interesting phenomenon is that, when $N_V$ is larger than $N_C$, no matter the semiconductor is n-type or p-type initially, it will always turn into n-type under extremely strong illuminations. Similarly, when $N_C$ is larger than $N_V$, the semiconductor will always be p-type under extremely strong illuminations.

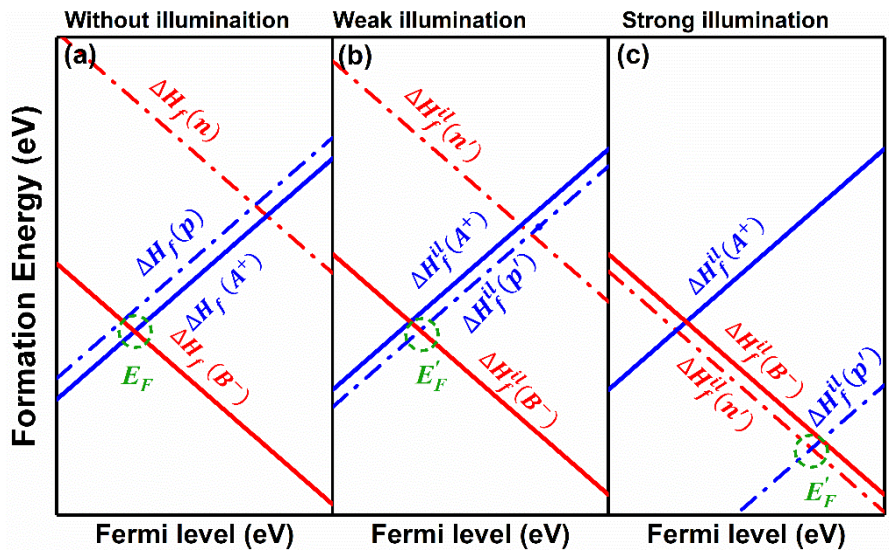

Figure 4. Diagrams to show formation energies of defects and band edge 'defects' as functions of the Fermi level under different illumination conditions. (a)-(c) Defect formation energies under no, weak, and strong illuminations, respectively. The solid lines are for defects and the dashed lines are for the band edge excitations.

## IV. Illumination effects on GaN:Mg

We now apply our defect theory under illuminations to study illumination effects on defect properties of GaN during the growth process. It is reported that $V_N$ is the main compensating center for the acceptor $Mg_{Ga}$ in GaN under the Ga-rich condition[30, 31]. Many literatures have reported that illumination can increase the hole concentration but the mechanisms are not clear[32-34].

By using the first-principles defect calculations (see the calculation details in the Supplemental Materials), we obtain that, $Mg_{Ga}$ has a (0/-) transition energy level around 0.25 eV above the VBM, while (+/0) and (3+/0) transition energy levels of $V_N$ locate at about 3.30 and 1.42 eV above the VBM, respectively, which are in good agreement with previous works [9, 35]. The calculated defect formation energies as functions of the Fermi level under the Ga-rich condition are shown in Fig. 5a. The lines for band edge defects are also given at $T$ = 1275 K, which is a typical experimental growth temperature. To obtain $N_C$ and $N_V$, the electron and hole effective masses of $0.2m_0$ and $1.5m_0$ are used [36], respectively. When illumination is applied, the effective temperature of electrons starts to increase from $T$ to $T'$. Note that, given illumination strengths and thus $T'$, defect concentrations, carrier densities, and the $E_F'$ can be obtained by self-consistently solving Eqs. (4)-(9) at a given growth temperature. In the followings, we use $\Delta n = n' - n_0$ to denote illumination strengths, where $n'$ and $n_0$ are the electron densities at electron temperatures of $T'$ and $T$ (here $T$=1275 K), respectively.

Without illumination, the Fermi level is pinned at about 0.97 eV above the VBM, because the defect excitation is dominate over the band-edge excitations [27]. When illumination is applied, the dashed lines in Fig. 5a will shift. Under weak illuminations when defect excitation is always much stronger than band edge excitations, the $E_F'$ doesn't change and the concentrations of all the charged defects will decrease because of increased illumination strengths and thus increased $T'$ according to our above analysis. Indeed, our results in Fig. 5b confirm our picture, i.e., both the concentrations of domination defects $Mg_{Ga}^-$ and the compensation defects $V_N^+$ decrease first until $\Delta n = 10^{11}$ cm$^{-3}$. When $\Delta n > 10^{11}$ cm$^{-3}$, the hole excitation starts to be comparable to or dominate over the excitation of $V_N^+$ (see Fig. 5b and Fig. S3b) and in this case, the $E_F'$ would be mainly determined by the concentrations of $Mg_{Ga}^-$ and holes. With the increase of photogenerated carriers, the $E_F'$ will move to the middle of the gap. According to Eq. (6), for positively charged defect $V_N^+$, because $q(E_t^\alpha - E_F')$ decreases and $T'$ increases, its concentration decreases. On the other hand, for negatively charged defect $Mg_{Ga}^-$, because $q(E_t^\alpha - E_F')$ increases faster than $T'$, its concentration slightly increases according to our calculations (see Fig. 5b and Fig. S3b). Nevertheless, our results show that the concentration of $Mg_{Ga}^-$ is nearly unaffected by illumination, but that of $V_N^+$ is significantly reduced under experimental growth conditions with an illumination strength corresponding to $\Delta n$ being about $10^{12}$-$10^{14}$ cm$^{-3}$ [37]. We find that the experiments only reported suppressed formation of $V_N$ but no change of Mg doping concentration [37, 38], which is in good agreement with our results. When the illumination further increases, i.e., $\Delta n > 10^{18}$ cm$^{-3}$ (suppose this can be achieved), the band edge excitations will surpass defect excitations (see Fig. 5b and Fig. S3c). In this case, we find that the $E_F'$ starts to be closer to the CBM, turning GaN from p-type to n-type. In addition, $q(E_t^\alpha - E_F')$ increases slower than $T'$ and therefore the concentration of $Mg_{Ga}^-$ decreases again.

Because of illumination effects on charged defects, it offers a novel way for manipulating defect formation during the growth process. For practical application, a sample is grown at a high temperature with appropriate illumination applied during the growth or post-growth process until a quasi-equilibrium status of electrons is achieved. Then the illumination is removed and the sample is quenched to room temperature, assuming defects are fixed and only electrons are redistributed [39]. Our simulation results for GaN are shown in Fig. 5c. As we can see, compared to the case without illumination during growth, samples grown with illumination have significantly shallower Fermi level close to the VBM, indicating a higher hole density is achieved. Our simulation results agree very well with experimental measurements[32, 40].

We also apply our theory to the Sb doping in CdTe and reasonably good agreement with available experiments is achieved. Especially, our theory can successfully explain the solubility increase of Sb doping due to illuminations during growth (see the Supplemental Materials), which can't be understood from previous theories. In addition to manipulating defect properties during the growth process, illumination during operations at room temperature can also have significant effects by enhancing band excitations, changing the ratio between charged and neutral defects and shifting the $E_F'$. We expect our theory to provide both quantitative and qualitative explanations for many interesting experimental phenomena under steady illuminations.

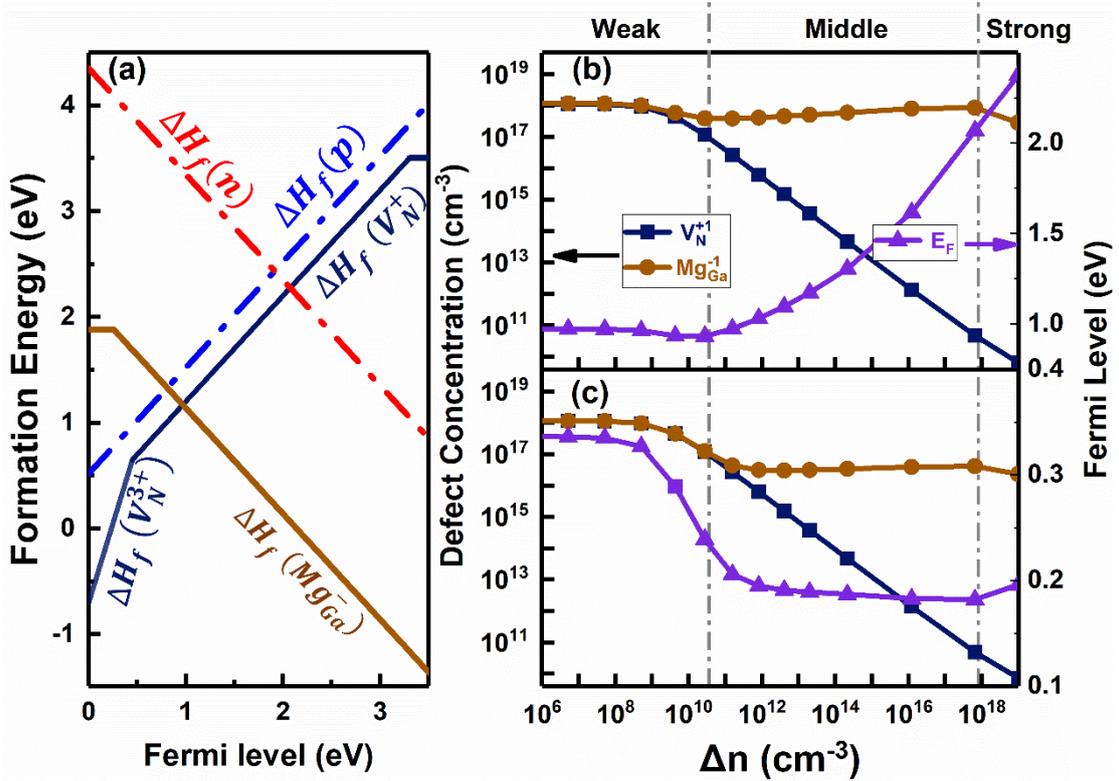

Figure 5. Illumination effects on defect properties of Mg doped GaN. (a) Defect formation energies as functions of the Fermi level in Mg doped GaN. (b) The concentrations of charged defects and effective Fermi level as functions of illumination strengths when GaN is grown at 1275K. (c) The concentrations of charged defects and the Fermi levels after the samples in Fig. (b) are quenched to 300 K and the illuminations are removed.

## V. Conclusion

In summary, we have proposed a self-consistent method to simulate the continuous and steady illumination conditions. We have proved that the illumination effects on formation energies of neutral defect and defect transition energy levels are negligible. To characterize electron distributions in a homogenous semiconductor under continuous and steady illuminations, we have pointed out that thermal excitations are equivalent to optical excitations for reaching a steady electron distribution. Therefore, the electron distribution can be characterized by using just one effective temperature $T'$ and one universal Fermi level $E_F'$. Based on the above concepts, we have uncovered the mechanisms of the illumination effects on defects by treating the band edge states explicitly in the same footing as the defect states. We have found that the formation energies of band edge 'defect' states shift with increased $T'$ of electrons, thus affecting the $E_F'$ of the total system, changing the ionic probabilities of defect states, and affecting concentrations of charged defects. Our proposed picture falls in line with the experimental observations and has been exemplified by GaN:Mg and CdTe:Sb systems. More experimental works are strongly called for to test our defect theory under illuminations.

# Acknowledgement

This work was supported in part by National Natural Science Foundation of China (Grant No. 12188101, 11991061, 61904035, 11974078). Computations were performed at the High-Performance Computing Center of Fudan University.